%% file: paper.tex
\documentclass[sigconf,authorversion,nonacm]{acmart}

\AtBeginDocument{%
  \providecommand\BibTeX{{%
    \normalfont B\kern-0.5em{\scshape i\kern-0.25em b}\kern-0.8em\TeX}}}

\setcopyright{acmcopyright}
\copyrightyear{2026}
\acmYear{2026}

\acmConference[TScIT 45]{45$^{th}$ Twente Student Conference on IT}{July 3,
  2026}{Enschede, The Netherlands}

\usepackage[htt]{hyphenat}
\usepackage{graphicx}
\graphicspath{{images/}}
\usepackage[most]{tcolorbox}
\usepackage{xcolor}
\usepackage{soul}
\definecolor{kg}{HTML}{58A92D}
\definecolor{question}{HTML}{010F6C}

\definecolor{input}{HTML}{E8F0FF}
\definecolor{sample}{HTML}{F3E9FE}
\definecolor{annotation}{HTML}{FAEDED}
\definecolor{generation}{HTML}{E9F6DF}
\definecolor{embedding}{HTML}{DAFEEE}

\def\Dataset{\texttt{ARLtR}}
\def\Framework{\texttt{ARLtR}}
\newcommand{\dimension}[3]{%
\begin{tcolorbox}[
    colback=#2!8,
    colframe=#2!80!black,
    boxrule=0pt,
    leftrule=1.5mm,
    sharp corners,
    left=3pt,
    right=3pt,
    top=2pt,
    bottom=2pt,
    boxsep=1pt
]
\textbf{#1}\par
\texttt{#3}
\end{tcolorbox}
}

\newcommand{\dimensiont}[3]{%
\begin{tcolorbox}[
    colback=#2!8,
    colframe=#2!80!black,
    boxrule=0pt,
    leftrule=1.5mm,
    sharp corners,
    left=3pt,
    right=3pt,
    top=2pt,
    bottom=2pt,
    boxsep=1pt
]
\textbf{#1}\par
#3
\end{tcolorbox}
}

\begin{document}

\title{All Relations Lead to Rome: Automated Knowledge Graph Creation and Question Generation}


\author{Matthijs Jansen op de Haar}
\email{m.m.t.p.jansenopdehaar@student.utwente.nl}
\affiliation{%
  \institution{University of Twente}
  \city{Enschede}
  \country{The Netherlands}
}

\author{Tobias Stähle}
\email{tobias.staehle@inf.ethz.ch}
\affiliation{%
  \institution{ETH Zürich}
  \city{Zürich}
  \country{Switzerland}
}

\author{Lorenzo Gatti}
\email{l.gatti@utwente.nl}
\affiliation{%
  \institution{University of Twente}
  \city{Enschede}
  \country{The Netherlands}
}

\renewcommand{\shortauthors}{Matthijs Jansen op de Haar}

\begin{abstract}
Large language models have substantially improved information retrieval and question answering; however, existing datasets generally support either vector-based retrieval over unstructured text or reasoning over knowledge graphs, without providing a unified representation that combines both paradigms. Moreover, current benchmarks rarely provide ground-truth entities, relations, and fact-grounded question-answer pairs aligned with the underlying corpus. To address this gap, we introduce \textbf{All Relations Lead to Rome (\Dataset)}, a unified framework for automated knowledge graph construction and fact-grounded question-answer generation. \Dataset\ jointly constructs a knowledge graph, embeddings, and question-answer pairs that are explicitly grounded in extracted entities, relations, and supporting textual evidence. We further instantiate the framework as a historical dataset centered on the Roman Empire, comprising over 19,000 entities, 16,000 chunks, and 8,400 question-answer pairs\footnote{https://huggingface.co/datasets/FaynePro/all-relations-lead-to-rome}. By tightly coupling symbolic graph representations with dense retrieval representations, \Dataset\ facilitates the evaluation and development of hybrid retrieval systems and semantic steering approaches within a single coherent resource.

\end{abstract}




\settopmatter{printacmref=false}

\begin{teaserfigure}
  \includegraphics[width=\textwidth]{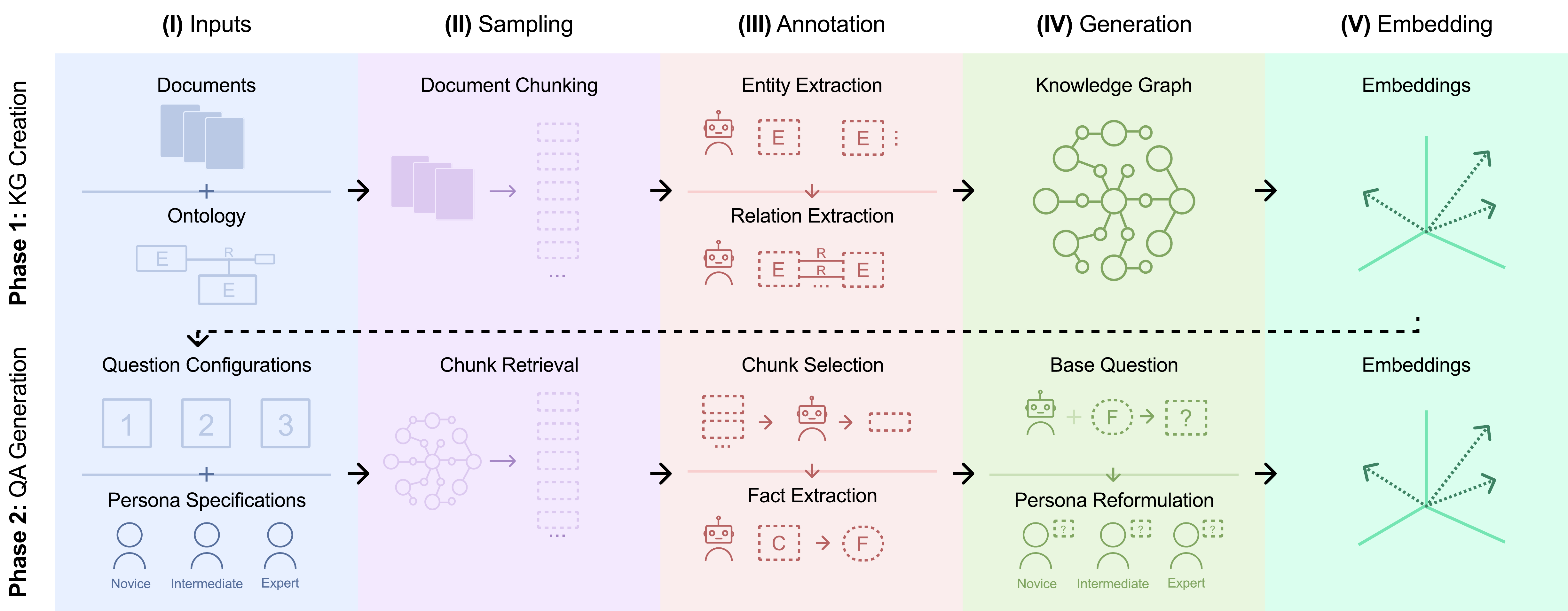}
  \caption{\normalfont In this work, we propose a framework for \textbf{(i) knowledge graph construction} and \textbf{(ii) fact-grounded question-answer generation}. \textbf{(i)} In the first phase, a corpus of documents is defined together with a \textit{weak} ontology specifying entity and relation types. The corpus is then segmented into chunks, after which the ontology guides entity extraction from each chunk. Relations between extracted entities are then identified, and the resulting structured information is integrated into a knowledge graph. Finally, embeddings are computed for both chunks and entities to support downstream retrieval. \textbf{(ii)} In the second phase, question configurations are defined together with persona specifications used for question formulation. Chunks are retrieved from the constructed knowledge graph based on these configurations. An LLM selects a relevant chunk and extracts a fact, which is used to generate an initial question. This question is then rewritten according to the specified personas. Finally, embeddings are computed for each generated question to enable comparison and retrieval.}
  \label{fig:teaser}
\end{teaserfigure}

\maketitle

\input{sections/01_introduction}
\input{sections/02_background}
\input{sections/03_framework}
\input{sections/04_dataset}
\input{sections/05_discussion}

\input{sections/06_conclusion}



\bibliographystyle{ACM-Reference-Format}
\bibliography{references}

\clearpage

\appendix

\section{Persona Descriptions and Prompt}
In this section, we present the persona descriptions used in \Dataset\ for reformulating each base question. These descriptions, along with their descriptors (e.g., Novice), are passed to an LLM to reformulate the base question.

\subsection{Persona Descriptions}

\dimension{Novice:}{white}{You are an inexperienced user with limited knowledge of the Roman Empire and no experience with RAG systems. You may use informal language, vague references, abbreviations, or incomplete terminology. Your questions might be more brief, or alternatively, too verbose without relevant additional context. Your questions should still convey the original intent, but may be less precise and may omit important contextual details, such as dates or use abbreviated terms and names.}

\dimension{Intermediate:}{gray}{You are a user with moderate knowledge of the Roman Empire. You are familiar with common historical figures, events, and concepts, but are far from a specialist. Your response can be of varying length, mostly containing relevant context, but occasionally throwing in unnecessary context. Your questions should generally be of moderate quality and understandable, on occasion using historical terminology, while preserving the original intent of the question, and sometimes omitting relevant details.}

\dimension{Expert:}{black}{You are a domain expert on the Roman Empire with extensive historical knowledge. You are comfortable using precise terminology, formal names, historical context, and specialized vocabulary. Your questions should be highly specific or detailed, unambiguous, and technically accurate while preserving the exact semantic meaning of the original question.}

\vspace{60mm}

\subsection{Persona Prompt}

\dimensiont{Persona Reformulation:}{white}{CONTEXT:
You are a system that rewrites questions from the perspective of a specific user role without changing their semantic meaning.

All questions concern the Roman Empire or related topics. The assigned user role defines the perspective and expertise from which the question should be phrased.

These questions are intended for a RAG system to retrieve relevant information.

ROLE:
You are a \{ persona \} user.

ROLE DESCRIPTION:
\{ description \}

TASK:
1. Rewrite the question provided under 'QUESTION TO REWRITE' without changing its semantic meaning.
2. Ensure the rewritten question reflects the assigned user role and role description.
3. Return only a JSON object. Do not include any additional text, explanations, or formatting.

QUESTION TO REWRITE:
\{ base question \}

OUTPUT FORMAT:
\{ "question": "..." \}}

\end{document}

%% file: sections/01_introduction.tex
\section{Introduction}
Large language models (LLMs) have significantly advanced the state-of-the-art in information retrieval (IR) and question answering, yet their performance remains highly dependent on the structure and quality of the underlying knowledge bases and data representations \cite{llms_in_ir, quality_necessity_of_kbs}. In particular, modern IR approaches such as Retrieval Augmented Generation (RAG) \cite{rag} increasingly rely on two complementary paradigms: vector-based retrieval over unstructured text chunks and structured reasoning over knowledge graphs (KGs) \cite{survey_vectors_to_kgs}. While both approaches have demonstrated strong performance on their own, they are rarely supported by a single unified dataset that enables reasoning across the two representations \cite{survey_vectors_to_kgs}.

On the one hand, vector-based retrieval systems, as commonly used in RAG, operate on chunked text representations embedded in a high-dimensional vector space \cite{rag}. These systems are effective at retrieving semantically relevant context, but they typically treat documents as isolated passages without explicit structure or relational grounding \cite{graphrag}. On the other hand, KG-based approaches encode information in terms of entities and relations, enabling structured querying and multi-hop reasoning \cite{graphrag, quality_necessity_of_kbs}. However, KG representations, on their own, can be detached from raw textual sources and are not always aligned with embedding-based representations at the chunk level, and vice versa \cite{noise_in_kg, rag, graphrag}. This separation leads to a fundamental limitation in existing question-answering (QA) datasets. Most QA datasets or benchmarks are designed either for unstructured retrieval or for structured reasoning and, in many cases, do not include ground-truth semantics (i.e., entities or relations) that can be used during retrieval to create selection spaces or to facilitate semantic steering (e.g., through entities) \cite{qa_survey,gap_kg_vs_vector,gap_kg_vs_vector_2}. The use of ground-truth semantics necessitates the inclusion of a knowledge base alongside unstructured chunks to facilitate selections (e.g., using categories), which is omitted in state-of-the-art QA approaches such as HotpotQA \cite{hotpotqa}. Consequently, this limits the ability to design or evaluate systems that require simultaneous reasoning over relational structure and dense semantic representations, as well as controlled retrieval guided by grounded semantics.

To address these gaps, we introduce \textbf{All Relations Lead to Rome} (\Dataset), a unified entity- and relation-centric framework that integrates vector-based retrieval, KG structure, and fact-grounded question-answer generation. Building on this framework, we additionally provide an accompanying dataset in the historical domain, also dubbed \Dataset. Each question is explicitly grounded in entities and relations extracted from a structured KG and linked to chunk-level representations that support dense retrieval. This dual alignment enables both symbolic reasoning and vector-based retrieval to operate on the same underlying resource, thereby bridging a key gap in current QA and retrieval datasets.

%% file: sections/02_background.tex
\section{Background}
KGs have become a central representation for structuring textual information, such as chunks, into entities and relations \cite{kg}. A KG typically encodes knowledge as triples, where entities are connected by typed relations, enabling both semantic reasoning and structured retrieval. The construction of such graphs is commonly performed via a pipeline comprising document preprocessing, entity extraction, and relation extraction, often guided by an ontology that defines the expected schema for entity and relation types \cite{kg,kg_survey,automated_kg_construction}. Depending on the setting, this ontology can be fully specified or weakly defined, allowing for more flexible extraction across domain-specific or open-domain corpora \cite{automated_kg_construction}. The resulting graph structure provides a compact representation of the underlying text while preserving explicit semantic relationships \cite{kg}.

Question-answer generation has been widely studied as a means of producing training and validation data for applications such as IR \cite{hotpotqa, qa_survey,qa_survey_generation}. QA generation systems typically operate by first identifying \textit{facts} from a corpus and then transforming them into natural-language question-and-answer pairs \cite{qa_survey}. More advanced approaches further condition question generation on additional variables such as difficulty level, context constraints, or persona specifications, enabling controlled generation of diverse question sets \cite{hotpotqa,qa_survey_generation,persona}. In fact-grounded settings, the key requirement is that each generated question can be traced back to an explicit supporting fact in the corpus, ensuring verifiability and reducing hallucination in downstream models trained on the data \cite{hotpotqa, qa_survey, qa_survey_generation}. Both knowledge graphs and QA generation have been extensively used in information retrieval systems. KGs enable structured retrieval by allowing systems to query over entities and relations, while also supporting hybrid retrieval strategies when combined with vector-based embeddings. In modern RAG systems, unstructured text is typically split into chunks, embedded in a vector space, and retrieved using dense retrieval \cite{rag}. These chunks are then used as context for LLMs, which are used as evidence for responses \cite{rag}. In other pipelines, structured knowledge bases such as KGs are combined with dense retrieval to improve both precision and interpretability, allowing systems to retrieve not only relevant passages but also explicitly related entities and relational context \cite{graphrag}.

\begin{figure}[h]
    \centering
    \includegraphics[width=0.59\linewidth]{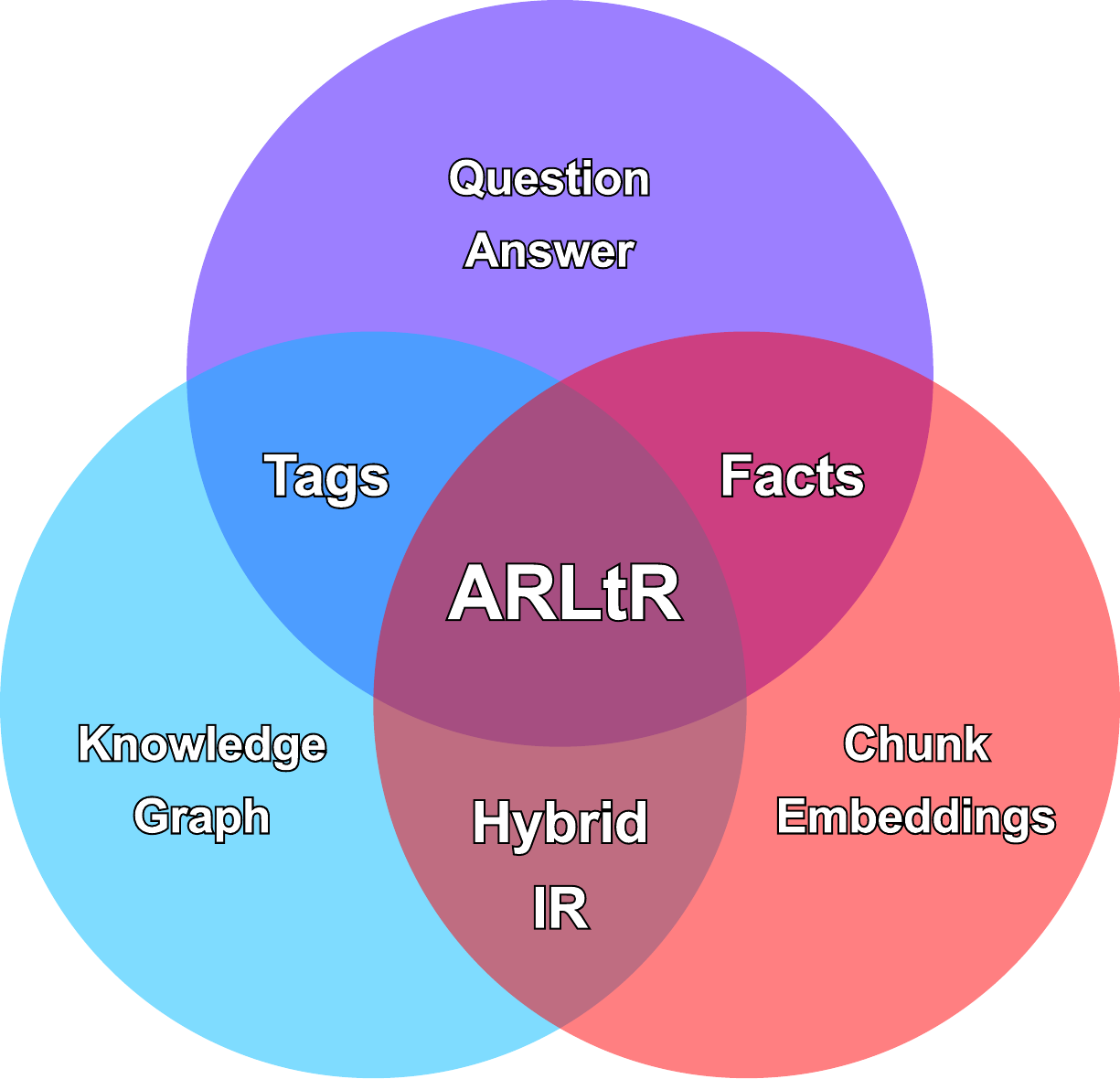}
    \caption{\normalfont \Dataset\ facilitates QA in both a knowledge graph and vector embeddings. Resulting in ground-truth facts, or answers, that coincide with ground-truth tags, or annotations.}
    \label{fig:venn_diagram}
\end{figure}

Despite these advances, a gap remains in existing methodologies for dataset construction. No existing dataset jointly provides \textbf{(i)} vector-based querying over embedded chunks, \textbf{(ii)} an explicit knowledge graph structure linking entities and relations, \textbf{(iii)} ground-truth entity and relation annotations (i.e., tags), and \textbf{(iv)} question-answer pairs that are explicitly grounded in factual content extracted from the same underlying corpus \cite{qa_survey,qa_survey_generation,gap_kg_vs_vector,gap_kg_vs_vector_2}. Existing resources typically address only subsets of these components, such as purely vector-based retrieval datasets or KG-centric benchmarks without aligned QA pairs \cite{gap_kg_vs_vector,gap_kg_vs_vector_2}. This limits the ability to evaluate or train systems that combine hybrid retrieval or alternatively use selection spaces and semantic steering to influence the IR process. In this work, we address this gap by proposing \Dataset, a unified framework that simultaneously constructs all four components in a consistent and fact-grounded manner,  illustrated in Figure \ref{fig:venn_diagram}

\section{Related Work}
In this section, we address QA datasets and frameworks related to \Dataset. In particular, we address the bottom segments as presented in Figure \ref{fig:venn_diagram}, namely \textbf{(i)} Vector-Based QA, \textbf{(ii)} Knowledge Graph QA, and \textbf{(iii)} Hybrid QA. 

\textbf{Vector-based QA} datasets have become central to the development of RAG systems, using dense retrieval over large unstructured corpora. Prominent benchmarks such as Natural Questions \cite{natural_questions}, TriviaQA \cite{triviaqa}, and HotpotQA \cite{hotpotqa} provide question-answer pairs aligned with chunks, supporting vector-based retrieval. These datasets are typically derived from unstructured text corpora such as Wikipedia\footnote{https://www.wikipedia.org/} and are designed to evaluate chunk retrieval and answer-generation pipelines. However, they largely operate in a purely unstructured setting, without explicit modeling of underlying semantics, such as entities and relations.

\textbf{Knowledge graph QA} datasets, in contrast, focus on structured reasoning over explicit relational representations. Benchmarks such as WebQuestionsSP \cite{webquestionssp}, ComplexWebQuestions \cite{complex_web_questions}, and MetaQA \cite{metaqa} ground questions in triples derived from knowledge graphs such as Wikidata\footnote{https://www.wikidata.org/} or DBpedia\footnote{https://www.dbpedia.org/}. These datasets emphasize compositional and multi-hop reasoning over entities and relations, often requiring graph traversal. This enables precise symbolic reasoning; however, they typically abstract away from the rich context found in chunk embeddings.

\textbf{Hybrid QA} datasets that combine vector-based retrieval with knowledge graph structure remain comparatively underexplored \cite{gap_kg_vs_vector,gap_kg_vs_vector_2,qa_survey,qa_survey_generation}. Recent efforts, such as KILT \cite{kilt}, partially bridge this gap by aligning textual evidence with structured semantics, yet they do not fully unify dense retrieval and relational information within a single coherent framework. As a result, few benchmarks support simultaneous reasoning over entities and relations, as well as vector-based retrieval over aligned textual representations. We address this gap in the literature by introducing \Dataset.

%% file: sections/03_framework.tex
\section{Framework and Methodology}
In this section, we illustrate the various phases and stages as described in Figure \ref{fig:teaser}. The phases include: \textbf{(i)} knowledge graph creation and \textbf{(ii)} question-answer generation. Moreover, we include a running example that corresponds to the implementation in the \Dataset\ dataset. 

\subsection{Knowledge Graph Creation}
In the knowledge graph creation phase, a given unstructured text dataset is split into chunks, after which entities are extracted using an LLM according to an ontology. These are then added to a knowledge graph, after which embeddings are computed for both entities and chunks to enable vector retrieval. 

\subsubsection{Inputs}
In the input stage, a set of unstructured textual documents is specified alongside an ontology. The ontology defines the allowed entity types and relations and can be either strongly or weakly defined; this is enforced during the \textit{annotation} stage. It is important to note that, unlike standard ontologies \cite{ontology}, the ontology is encoded in a textual representation and generally \textit{weakly} defined (i.e., only entity and relation types). This will allow an LLM to enforce the ontology through instructions provided in a system prompt. Nevertheless, it is possible to enforce an ontology after extraction, as entities and relations can be selectively added to a KG. In the \Dataset\ dataset, a list of entities and relation types is specified. 

\dimensiont{Example:}{input}{A set of \texttt{HTML} pages and \texttt{PDF} documents is used that contain relevant information about the \textit{Roman Empire}. Furthermore, an ontology is defined that includes entity types \texttt{PERSON} and \texttt{LOCATION}, and relation type \texttt{LIVES}.}

\subsubsection{Sampling}
In the sampling stage, the documents are chunked into chunks, or strings; these chunks have a fixed size (i.e., tokens) and also exhibit overlap with other chunks. The latter is to ensure that relevant context is not lost during the chunking process, in which sentences or meaning are split \cite{rag, graphrag}. The right balance of chunk length is a fundamental design decision in creating a knowledge graph, as larger chunks will require fewer LLM calls during annotation, while smaller chunks can improve recall of information \cite{graphrag, chunk_size}. In the design of the \Framework\ dataset, medium-sized chunks of 100 tokens were used. However, overlap was omitted due to the small chunk size and to prevent over-representation of facts in the QA Generation stage. 

\dimensiont{Example:}{sample}{Based on the provided \texttt{HTML} pages and \texttt{PDF} documents, each is chunked into 600-token chunks with 100-token overlap between chunks. This result strikes a good balance between annotation efficiency and the avoidance of context loss across poorly split documents.}

\subsubsection{Annotation}
After the corpus has been chunked, the annotation stage is applied using an LLM-based entity tagger, a method commonly used in knowledge graph generation pipelines \cite{automated_kg_construction,graphrag}. In this stage, entities can be extracted and used directly as metadata within a standard vector database. However, for knowledge graph construction, additional relational information is required. This can be obtained either by directly extracting triplets containing entities and relations, or by separating the process into entity annotation followed by relation extraction over the identified entities. The latter approach follows a \textit{chain of thought} \cite{chain_of_thought}, in which relations are inferred from previously extracted entities. While this increases the number of LLM calls, it improves annotation accuracy, particularly in settings with larger or more complex ontologies \cite{chain_of_thought}. In constructing the \Dataset\ dataset, we adopt this two-step process for triplet identification.

\dimensiont{Example:}{annotation}{Using the specified ontology, a chunk is tagged by an LLM-based entity tagger, extracting entities (e.g. \texttt{Augustus}) according to the specified entity types. These are subsequently used to extract relations, using both the chunk and specified entities (e.g., \texttt{Augustus $\rightarrow$ ruled $\rightarrow$ Roman Empire}). Using a \textit{chain of thought} ensures better knowledge graph annotation quality.}

\subsubsection{Generation}
During the knowledge graph generation stage, extracted entities and relations are integrated into the knowledge graph. Entities are merged based on their title, regardless of capitalization, to avoid duplicate entries when an entity with the same name already exists. Subsequently, relations are added using the same merging strategy to ensure consistency at the graph level. While this reduces redundancy in the constructed dataset, its effectiveness depends on the underlying domain and naming conventions, particularly when distinct entities share the same name. Furthermore, this stage should enforce that entities and relations conform to valid types, as LLM-based annotation may introduce incorrect or inconsistent labels \cite{hallucinations,tagging_inaccuracies}. In practice, annotations can be added in batches or one by one during knowledge graph creation.

\dimensiont{Example:}{generation}{Each of the annotated entities (e.g., \texttt{Augustus}), and relations (e.g., \texttt{LIVES}) is added one-by-one into the dataset. Entities with different capitalization are merged upon insertion to the knowledge graph (e.g., \texttt{augustus} or \texttt{AUGUSTUS}.}

\subsubsection{Embedding}
In the embedding stage, vector representations are computed for each chunk using a pre-trained embedding model. These embeddings are defined in a fixed-dimensional vector space, where higher dimensionality generally allows for a richer representation of semantic information. The resulting vectors are stored as attributes of the corresponding chunks and can be leveraged for dense retrieval methods, such as cosine similarity. In addition, embeddings can also be computed for entities, enabling entity-level matching and comparison via dense retrieval. In the \Dataset\ dataset, both chunk-level and entity-level embeddings are computed and stored within the knowledge graph to support retrieval tasks.

\dimensiont{Example:}{embedding}{Given some chunk, embeddings are computed using a pre-trained embedding model. These are then added as an attribute to the knowledge graph. Moreover, to facilitate dense retrieval, entities and their descriptions have been turned into embeddings.}

\subsection{Question-Answer Generation}
In the question-answer generation phase, question configurations and persona specifications are pre-defined and guide the subsequent process. Based on this configuration, relevant entities are selected, and associated chunks are then retrieved to ensure contextual relevance. An LLM-based fact extractor then selects one or multiple facts from the retrieved chunks. Using these extracted facts, a base question is formulated, where the facts serve as the ground-truth answer. This base question is subsequently rewritten according to the specified persona configurations. Finally, embeddings are computed over the generated questions to support retrieval.

\subsubsection{Inputs}
During the input stage, question configurations are specified, defining and constraining the subsequent stages. These configurations are summarized in Table \ref{tab:question_configurations}, but can be extended depending on the target domain and specific QA use cases. Based on this configuration, different strategies for chunk retrieval, selection, and question generation can be applied. In addition, persona specifications are defined to describe user profiles and their corresponding levels of expertise. These personas are subsequently used in the persona reformulation stage to guide the generation of tailored question variants.

\begin{table}[t]
\caption{\normalfont Question Configurations in \Dataset}
\label{tab:question_configurations}
\centering
\begin{tabular}{l l l}
\toprule
\textbf{Category}&\textbf{Type}&\textbf{Description} \\
\midrule
Question Type & Factoid & Requiring a single fact\\
&Open& Requiring multiple facts\\
Relational& Entity& Regarding an entity\\
&Relation& Regarding a relation\\
Complexity & Single & Single entity, or relation\\
&Double& Two entities, or neighbor\\
Persona&Novice& Lowest user-experience\\
&Intermediate& Average user-experience\\
&Expert& Advanced user-experience\\
Existent&In Dataset& Supported by facts\\
&Not in Dataset&Unsupported by facts\\
\bottomrule
\end{tabular}
\end{table}

The categories presented in Table \ref{tab:question_configurations} are grounded in the literature. The question type category distinguishes between factoid and open questions, which require one or multiple facts to answer, respectively. Factoid questions are typically used to assess whether a response is discretely correct, whereas open questions are evaluated based on their exhaustiveness \cite{natural_questions,factoid}. The relation category encodes whether a question concerns a single entity or one of its relations. Alternatively, questions may involve neighboring entities, requiring multi-hop reasoning \cite{relational_questions,relational_questions_two}. This aspect is captured by the complexity category, which differentiates between single-entity and two-entity configurations for entity-based questions. For relational questions, it determines whether the query targets a direct relation or a neighboring entity, the latter requiring an entity hop. Furthermore, the persona category encodes varying levels of user expertise and is commonly used in Human-Computer Interaction research to better model user diversity \cite{user_expertise}. Lastly, the existent category indicates whether a question-answer pair is supported by the dataset, thereby enabling the inclusion of unanswerable or misleading questions to evaluate system precision.

\dimensiont{Example:}{input}{Each question is constructed by selecting a question type (e.g., factoid or open), an entity (e.g., \texttt{Augustus}), and a relation (e.g., \texttt{RULED}); depending on the complexity setting, a neighbor entity (e.g., \texttt{Roman Empire}) may be added, which will result in a question being asked about said neighbor. (e.g., ``When was the empire ruled by Augustus founded?'') 
}

\subsubsection{Sampling}
\begin{figure}[h]
    \centering
    \includegraphics[width=1\linewidth]{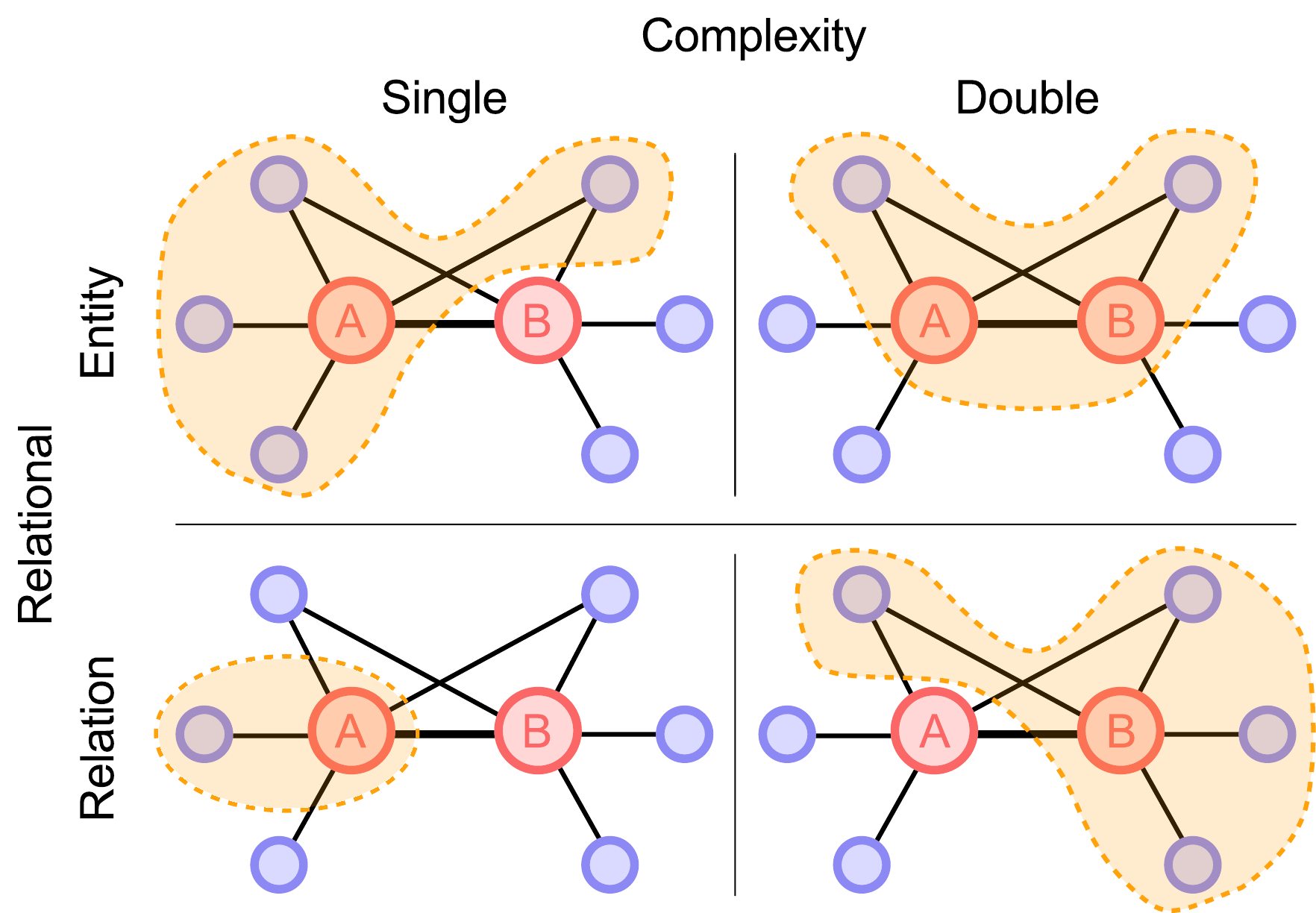}
    \caption{\normalfont Sampling Variations between Complexity and Relational Question Categories between Entity (A) and Neighbor (B)}
    \label{fig:complexity-relational}
\end{figure}

In the sampling stage, the provided question configuration is used to retrieve chunks for fact extraction. This is achieved by randomly selecting entities and relations associated with a sufficient number of chunks, ensuring they are meaningfully connected and can support factual extraction across the dataset. This process guarantees that extracted facts are grounded in the KG and remain aligned with the selected entities. The different sampling strategies are illustrated in Figure \ref{fig:complexity-relational}. Each configuration is defined as follows: \textbf{(i)} Single-Entity: all chunks associated with a given entity are considered relevant for fact extraction; \textbf{(ii)} Double-Entity: chunks are selected based on the intersection between two entities, capturing shared information relevant for extraction; \textbf{(iii)} Single-Relation: a specific relation type is selected for a given entity, and the corresponding chunks are used to extract a relevant fact; and \textbf{(iv)} Double-Relation: a relational hop is required, where chunks from a neighboring entity are included, and the intersection is used to extract the multi-hop relation. In \Dataset\, for each retrieval method, for a given configuration, the number of chunks should at least be five. This ensures that entities are always related and relevant, given the underlying dataset. 

\dimensiont{Example:}{sample}{Given some combination of question categories, an entity with sufficient chunks is randomly selected (e.g., \texttt{Augustus}). In this case, a \textit{Double-Entity} question was specified. Therefore, a neighboring entity that has a sufficient amount of chunks shared with \texttt{Augustus} is chosen, in this case \texttt{Rome}. The intersection of shared chunks is then sampled for the proceeding steps.}

\subsubsection{Annotation}
During the annotation stage, an LLM is given a set of chunks and tasked with extracting one or multiple facts. This fact extraction is framed from the question configuration, as different configurations require different facts. Each fact should correspond to the previously specified entity and possible relations, rather than being an unrelated fact from a given chunk. Doing the extraction prior to question generation also benefits from \textit{chain of thought} \cite{chain_of_thought}. 

\dimensiont{Example:}{annotation}{Given a set of chunks related to \texttt{Augustus} and the relation \texttt{LIVES}, the LLM is prompted to extract a fact aligned with this configuration. From chunks such as ``Augustus spent much of his reign in Rome,'' it extracts the fact \texttt{Augustus $\rightarrow$ lived in $\rightarrow$ Rome}. For a Double-Relation configuration, additional chunks may lead to a multi-hop fact (e.g., \texttt{Augustus $\rightarrow$ ruled $\rightarrow$ Roman Empire $\rightarrow$ capital $\rightarrow$ Rome}).}

\subsubsection{Generation}
Using the extracted facts, an LLM generates a base question that requires those facts to answer. Furthermore, the subjects (i.e., entities) are added to the question, enabling assessment of automated tagging strategies. This results in a question-answer pair, where the answer is equivalent to a fact that is aligned with a ground-truth entity. After the base question formulation is created, it is reformulated for each persona specification, resulting in formulation variations to facilitate comparisons across different user experience levels. These specifications should include a description of the persona (or user) and indicate whether they can provide additional details or leave a question ambiguous. The persona configurations used in \Dataset\ are specified in Table \ref{tab:personas}, and the descriptions and prompt are in Appendix A. 

\begin{table}[t]
\centering
\caption{Persona Expertise Levels and Base Descriptions}
\label{tab:personas}
\begin{tabular}{ll}
\toprule
\textbf{Personsa Lvl.} & \textbf{Description} \\
\midrule
Novice & Informal terminology and omitted context \\
Intermediate & Adequate terminology and base context \\
Expert & Precise terminology and additional context \\
\bottomrule
\end{tabular}
\end{table}

\dimensiont{Example:}{generation}{Given the extracted fact \texttt{Augustus $\rightarrow$ ruled $\rightarrow$ Roman Empire}, the LLM generates a base question such as ``Which empire did Augustus rule?''. The entity \texttt{Augustus} is explicitly included to enable evaluation of entity recognition. The answer corresponds directly to the fact (i.e., \texttt{Roman Empire}). This base question is then reformulated based on persona specifications; e.g., a novice may ask ``What place did Augustus rule?'', while an expert may ask ``Which political entity was governed by Augustus during his reign?'', producing multiple variations aligned with different user expertise levels.}

\subsubsection{Embedding}
During the final stage, embeddings are computed using a pre-trained embedding model applied to the persona-formulated questions. This will enable dense retrieval over the chunks in the KG, using methods such as \textit{cosine similarity}. These embeddings should be compatible with the ones presented in the KG and use the same embedding model.  

\dimensiont{Example:}{embedding}{The question ``Which empire did Augustus rule?'' is encoded using a pre-trained embedding model, producing a dense vector representation. This embedding can then be compared with chunk embeddings in the KG using cosine similarity to retrieve chunks that mention \texttt{Augustus} and the \texttt{Roman Empire}. By using the same embedding model as the KG, the retrieved chunks remain semantically aligned with the question for accurate dense retrieval.}

%% file: sections/04_dataset.tex
\section{Dataset}
In this section, we discuss the \Dataset\ dataset, which uses the \Dataset\ framework to create a KG and QA for historical documents related to the \textit{Roman Empire}. We provide our dataset as a knowledge graph, and both existent and non-existent question-answer pairs\footnote{https://huggingface.co/datasets/FaynePro/all-relations-lead-to-rome}.

\subsection{All Relations Lead to Rome}
All Relations Lead to Rome (\Dataset) is a KG and QA dataset with entities and relations about the \textit{Roman Empire}, made in Neo4j\footnote{https://neo4j.com/}. In the extraction and embeddings, we used an LLM (i.e., \textit{minimax 2.7}\footnote{https://www.minimax.io/models/text/m27}) and an embedding model (i.e., \textit{gemini-embedding-2}\footnote{https://deepmind.google/models/gemini/embedding/}) for the various stages as illustrated in Figure \ref{fig:teaser}. The input dataset comprises a representative sample of 300 articles from Wikipedia, spanning the \textit{Roman Empire} to the \textit{Byzantine Empire} and covering people, locations, and historical events. 

\begin{figure}[h]
    \centering
    \includegraphics[width=1\linewidth]{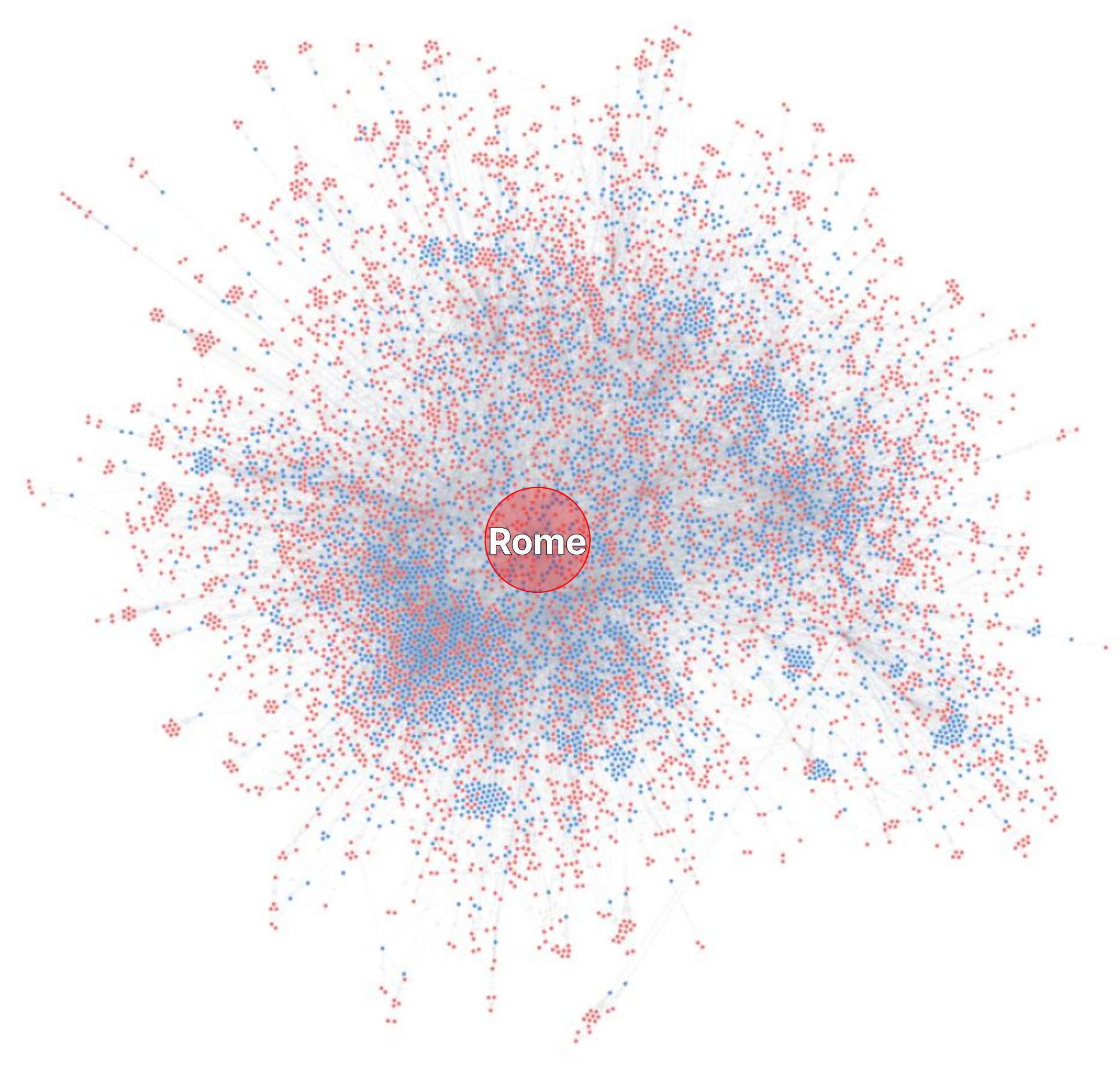}
    \caption{\normalfont Subset of the \Dataset\ KG, with Rome Highlighted}
    \label{fig:kg}
\end{figure}

\Dataset\ contains 16,069 chunks, each approximately 100 tokens, totaling \textbf{1.6M tokens}. All chunks are linked to at least one entity, this being the corresponding Wikipedia article, making for 19,374 total entities, ranging from people to locations. These entities have 25,304 total relations and are linked to various chunks 88,135 times. This means that each chunk contains an average of 4.55 entities. A subset of approximately 1000 nodes from the knowledge graph is shown in Figure \ref{fig:kg}, where the entity \textit{Rome} is highlighted as having the highest degree in the graph (i.e., 3,476 total links). Therefore, \textit{All Relations Lead to Rome}.

Furthermore, the knowledge includes embedding attributes for all chunks and entities, generated using the pre-trained \textit{gemini-embedding-2} model\footnotemark[7] (i.e., with 3,072 dimensions). This ensures that embeddings do not need to be recomputed for any future studies that use \Dataset. Moreover, the dataset contains vector indices for both \textit{entities} and \textit{chunks}, which is supported in Neo4j\footnotemark[5]. This allows for efficient dense retrieval over the graph without needing to compute embeddings one by one.   

\begin{figure}[t]
    \centering
    \includegraphics[width=1\linewidth]{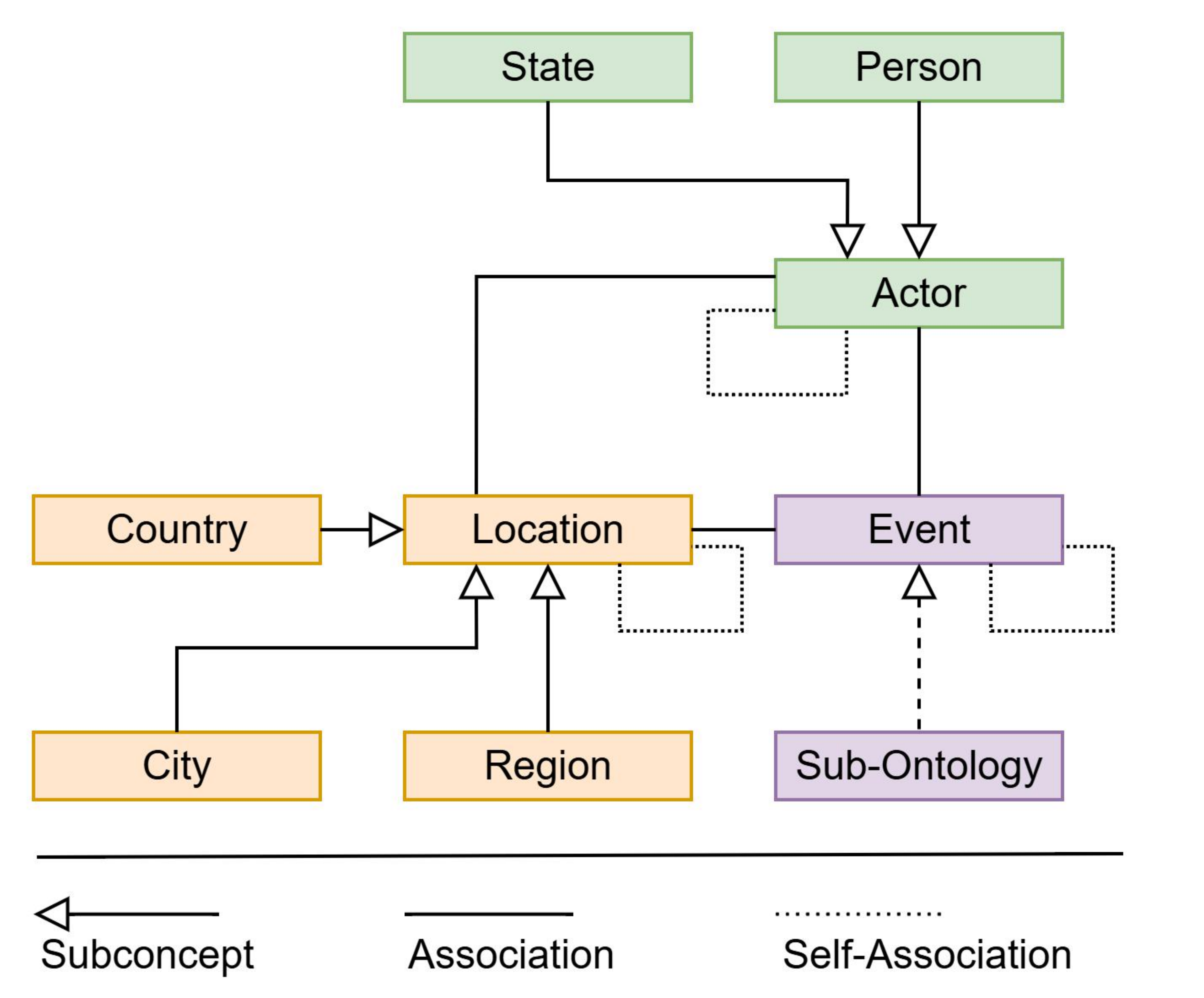}
    \caption{\normalfont Ontology Grounded in the Historical Domain}
    \label{fig:ontology}
\end{figure}

\subsection{Ontology}
In the design of \Dataset, we adopt an ontology grounded in existing historical and ontological literature \cite{ontology_one,ontology_two,ontology_three}. As discussed, we employ a \textit{weak} ontology to facilitate extraction, meaning that all defined entity types are allowed to relate to one another without hard constraints on admissible relation pairs. This ensures that the dataset is more aligned with KGs used in practice. In the historical domain, the primary categories are commonly identified as Actors, Locations, and Historical Events \cite{ontology_one,ontology_three}.

We further refine Actors into two subtypes: \texttt{PERSON} and \texttt{STATE}. For Locations, we distinguish between \texttt{CITY}, \texttt{COUNTRY}, and \texttt{REGION}. For Historical Events, while the literature differentiates between event classes such as wars, scientific discoveries, and religious events, we collapse these into a single unified category, \texttt{HISTORICAL EVENT}. This abstraction ensures that any historically salient event can be consistently identified and tagged by the LLM-based annotator, without requiring fine-grained event-type supervision. The resulting ontology is illustrated in Figure \ref{fig:ontology}, with all entity types summarized in the \textit{entity type} box. During the construction of \Dataset, we manually resolved ontological inconsistencies, including mislabeled entities and conflicting relation assignments \cite{graphrag,automated_kg_construction,kg_llm_creation}. Additionally, any articles that were not considered one of the aforementioned entity types were simply labeled as \texttt{ARTICLE}, as each entity has exactly one type. 

\dimension{Entity Types:}{black}{CITY, COUNTRY, REGION, HISTORICAL EVENT, PERSON, STATE, ARTICLE}

The relation types are further grounded in the literature \cite{ontology_one,ontology_two,ontology_three}. However, sources commonly diverge on the precise relation types, and often focus on specifying entities or specific triplets. Therefore, we created an exhaustive list that combines the specified relation types from the literature. In addition, by specifying relation types, the quality of the KG improves drastically, as the LLM-annotator will not include irrelevant relation types. The various relation types are specified in the \textit{relation types} box 

\dimension{Relation Types:}{black}{FOUNDED BY, LEADER OF, MAJOR EVENT IN, LOCATED IN, CAPITAL OF, INVOLVED IN, OCCURRED IN, MEMBER OF, ENEMY OF, ALLY OF, DIED IN, FOUGHT IN, MINOR EVENT IN, BORN IN, HAS CHUNK, MENTIONS}

It should be noted that for \texttt{MENTIONS} and \texttt{HAS CHUNK}, these indicate whether an entity is connected to a given chunk. This facilitates the retrieval of chunks and defines whether an entity appears or is connected to a given chunk.

\subsection{Question-and-Answers}
Alongside the knowledge graph, \Dataset\ also provides an extensive list of entity and fact-grounded question-and-answer pairs. A total of 8,400 QA pairs is provided. 6,000 of these questions are grounded in entities in the knowledge base, and the remaining 2,400 concern entities or facts not present in the knowledge base. It is important to note that some of the latter questions might have entities in the knowledge graph, but no facts. For example, \textit{United States} is included as an entity because it is mentioned in one of the articles, even though it is not relevant to the \textit{Roman Empire}. This is a natural result of using underlying documents for fact extraction. Nevertheless, facts or answers remain nonexistent in the dataset. This facilitates the evaluation of retrieval precision, as false positives may be retrieved for a question with no answer. Each possible question configuration from Table \ref{tab:question_configurations} is equally represented using 350 distinct entities (i.e., number of entities, or $\# Ent.$) in the dataset. Described with:

\begin{equation}
    QA_{pairs} = \#Ent. * \#Per.*\#Comp.*\#Rel.*\#Que.
\end{equation}

In the case of \Dataset\ this results in the following equation:

\begin{equation}
    QA_{pairs} = \#Ent. * 24
\end{equation}

%% file: sections/05_discussion.tex
\section{Discussion}

\subsection{General Implications}
We present \Dataset, a unified framework and dataset that facilitates both dense retrieval and knowledge graph construction in a tightly coupled manner. In addition, \Dataset\ enables the generation of QA-pairs that are explicitly grounded in textual chunks, extracted facts, and structured entity–relation representations derived from the underlying corpus. This joint grounding introduces a stronger form of supervision than text-only QA generation, as answers can be traced simultaneously to raw evidence and to symbolic graph structure. As a result, \Dataset\ provides a controlled environment for systematically comparing retrieval strategies, particularly retrieval-augmented generation methods such as base RAG \cite{rag} and graph-based extensions such as GraphRAG \cite{graphrag}, which are otherwise difficult to evaluate in current benchmarks.

Beyond evaluation, the availability of ground-truth entities and relations enables a range of structured selection and control mechanisms over the dataset. In particular, retrieval can be explicitly conditioned on entity sets or relation types, enabling semantic steering of the retrieval process, in which the high-level graph structure determines the most appropriate retrieval strategy for a given query. This introduces a more fine-grained interface between symbolic structure and neural retrieval, and enables hybrid systems that dynamically adapt retrieval pathways based on graph-aware semantics rather than relying solely on embedding similarity.

Furthermore, \Dataset\ provides direct utility for downstream system design and domain-specific analysis, particularly in the historical domain of the Roman Empire, where structured representations enable new forms of exploratory analysis, retrieval, and reasoning over interconnected entities and events. While the current instantiation is grounded in this domain, the underlying \Dataset\ framework is intentionally domain-agnostic. By adopting a weakly defined ontology, the framework reduces the need for extensive upfront ontological engineering, while still preserving sufficient structure for reliable entity and relation extraction. This design choice significantly lowers the barrier to constructing comparable knowledge graphs in other domains. Consequently, the approach generalizes naturally to settings such as the medical or financial domains, where structured knowledge representations are similarly valuable, yet rigid ontological design is often costly and difficult to maintain. Overall, \Dataset\ not only provides a dataset for benchmarking retrieval and generation systems but also serves as a reusable infrastructure for constructing, manipulating, and evaluating knowledge graphs and grounded QA datasets within a unified, extensible framework.

\subsection{Limitations and Future Work}

\Dataset\ makes use of automated LLM-based annotation in order to extract entities and relations from each chunk. This results in inherent inaccuracies in the tagging process when compared to human annotation, as extraction errors, missing relations, or incorrect entity boundaries may occur. While it is possible to incorporate human annotation within the proposed framework, this introduces a trade-off between annotation quality, time, and cost. Future work may explore hybrid approaches that combine LLM-based annotation with selective human verification to improve overall quality while maintaining scalability.

Furthermore, answer generation within \Dataset\ is not constrained to uniquely identifiable facts. In practice, a given question may be answerable using multiple valid facts or combinations of entities, resulting in ambiguity in the ground-truth answer space. While this introduces challenges for strict evaluation, it also reflects realistic retrieval settings in which multiple supporting facts may exist. As such, the dataset is better suited for evaluation under recall-oriented metrics, where systems are required to retrieve a set of appropriate answers rather than a single exact match. Future work may investigate mechanisms for clustering or normalizing equivalent answers to enable more fine-grained evaluation.

Finally, the \Dataset\ dataset is constructed in the historical domain, introducing domain-specific characteristics in both the knowledge graph and the underlying text. In particular, historical data often exhibits moderate graph density and well-defined entity relationships, which may not generalize to domains with either sparser or significantly denser relational structures. While \Dataset\ itself is of above-moderate size, the proposed framework is not limited to this setting and can be used to construct datasets of varying scale and complexity. This includes both smaller subsets of \Dataset\ that may serve as proxies for low-resource knowledge graphs and larger, domain-specific extensions. Future work should therefore explore the application of the framework across diverse domains and graph configurations to better understand its generalizability.

%% file: sections/06_conclusion.tex
\section{Conclusion}
In this work, we presented \Dataset, a unified framework and accompanying dataset that bridges the gap between vector-based retrieval and knowledge graph reasoning. Unlike existing resources that only provide subsets of these components, \Dataset\ jointly constructs chunk embeddings, structured entity-relation representations, ground-truth annotations, and fact-grounded question-answer pairs from a shared corpus. This unified design enables both symbolic reasoning and dense retrieval to operate over the same underlying resource, facilitating controlled evaluation of hybrid retrieval approaches and graph-enhanced retrieval-augmented generation systems. Furthermore, the use of a \textit{weak} ontology and automated construction pipeline lowers the barrier to creating comparable datasets in other domains. Consequently, \Dataset\ serves not only as a benchmark for retrieval and question-answering research but also as a reusable and extensible framework for constructing grounded knowledge graphs and hybrid retrieval resources across diverse application settings.